\documentclass[a4paper,fleqn,usenatbib]{mnras}

\usepackage{mathptmx}

\usepackage[T1]{fontenc}
\usepackage{ae,aecompl}
\usepackage{color}

\usepackage{graphicx}	
\usepackage{amsmath}	
\usepackage{amssymb}	
\usepackage{array}
\usepackage{threeparttable}

\newcolumntype{L}{>{\centering\arraybackslash}m{3.3cm}}

\title[]{MAGIC observations of the microquasar V404 Cygni during the 2015 outburst}

\author[M.~L.~Ahnen~et.~al.]{
M.~L.~Ahnen$^{1}$,
S.~Ansoldi$^{2,13}$,
L.~A.~Antonelli$^{3}$,
C.~Arcaro$^{4}$,
A.~Babi\'c$^{5}$,
B.~Banerjee$^{6}$,\newauthor
P.~Bangale$^{7}$,
U.~Barres de Almeida$^{7}$,
J.~A.~Barrio$^{8}$,
J.~Becerra Gonz\'alez$^{9,10}$,\newauthor
W.~Bednarek$^{11}$,
E.~Bernardini$^{12,25}$,
A.~Berti$^{2,26}$,
B.~Biasuzzi$^{2}$,
A.~Biland$^{1}$,
O.~Blanch$^{13}$,\newauthor
S.~Bonnefoy$^{8}$,
G.~Bonnoli$^{14}$,
R.~Carosi$^{14}$,
A.~Carosi$^{3}$,
A.~Chatterjee$^{6}$,
P.~Colin$^{7}$,\newauthor
E.~Colombo$^{9,10}$,
J.~L.~Contreras$^{8}$,
J.~Cortina$^{13}$,
S.~Covino$^{3}$,
P.~Cumani$^{13}$,
P.~Da Vela$^{14}$,\newauthor
F.~Dazzi$^{3}$,
A.~De Angelis$^{4}$,
B.~De Lotto$^{2}$,
E.~de O\~na Wilhelmi$^{15}$,
F.~Di Pierro$^{4}$,\newauthor
M.~Doert$^{16}$,
A.~Dom\'inguez$^{8}$,
D.~Dominis Prester$^{5}$,
D.~Dorner$^{17}$,
M.~Doro$^{4}$,
S.~Einecke$^{16}$,\newauthor
D.~Eisenacher Glawion$^{17}$,
D.~Elsaesser$^{16}$,
M.~Engelkemeier$^{16}$,
V.~Fallah Ramazani$^{18}$,\newauthor
A.~Fern\'andez-Barral$^{13}$\textcolor[rgb]{0.,0.,1.}{$^{\star}$},
D.~Fidalgo$^{8}$,
M.~V.~Fonseca$^{8}$,
L.~Font$^{19}$,
C.~Fruck$^{7}$,\newauthor
D.~Galindo$^{20}$,
R.~J.~Garc\'ia L\'opez$^{9,10}$,
M.~Garczarczyk$^{12}$,
M.~Gaug$^{19}$,
P.~Giammaria$^{3}$,\newauthor
N.~Godinovi\'c$^{5}$,
D.~Gora$^{12}$,
S.~Griffiths$^{13}$,
D.~Guberman$^{13}$,
D.~Hadasch$^{21}$,
A.~Hahn$^{7}$,\newauthor
T.~Hassan$^{13}$,
M.~Hayashida$^{21}$,
J.~Herrera$^{9,10}$,
J.~Hose$^{7}$,
D.~Hrupec$^{5}$,
G.~Hughes$^{1}$,\newauthor
K.~Ishio$^{7}$,
Y.~Konno$^{21}$,
H.~Kubo$^{21}$,
J.~Kushida$^{21}$,
D.~Kuve\v{z}di\'c$^{5}$,
D.~Lelas$^{5}$,\newauthor
E.~Lindfors$^{18}$,
S.~Lombardi$^{3}$,
F.~Longo$^{2,26}$,
M.~L\'opez$^{8}$,
C.~Maggio$^{19}$,
P.~Majumdar$^{6}$,\newauthor
M.~Makariev$^{22}$,
G.~Maneva$^{22}$,
M.~Manganaro$^{9,10}$,
K.~Mannheim$^{17}$,
L.~Maraschi$^{3}$,\newauthor
M.~Mariotti$^{4}$,
M.~Mart\'inez$^{13}$,
D.~Mazin$^{7,21}$,
U.~Menzel$^{7}$,
M.~Minev$^{22}$,
R.~Mirzoyan$^{7}$,\newauthor
A.~Moralejo$^{13}$,
V.~Moreno$^{19}$,
E.~Moretti$^{7}$\thanks{Corresponding authors: E.~Moretti, email:  \href{mailto:moretti@mpp.mpg.de}{moretti@mpp.mpg.de}, A.~Fern\'andez-Barral, email:  \href{mailto:afernandez@ifae.es}{afernandez@ifae.es}},
V.~Neustroev$^{18}$,
A.~Niedzwiecki$^{11}$,\newauthor
M.~Nievas Rosillo$^{8}$,
K.~Nilsson$^{18,27}$,
D.~Ninci$^{13}$,
K.~Nishijima$^{21}$,
K.~Noda$^{13}$,\newauthor
L.~Nogu\'es$^{13}$,
S.~Paiano$^{4}$,
J.~Palacio$^{13}$,
D.~Paneque$^{7}$,
R.~Paoletti$^{14}$,
J.~M.~Paredes$^{20}$,\newauthor
X.~Paredes-Fortuny$^{20}$,
G.~Pedaletti$^{12}$,
M.~Peresano$^{2}$,
L.~Perri$^{3}$,
M.~Persic$^{2,3}$,\newauthor
P.~G.~Prada Moroni$^{23}$,
E.~Prandini$^{4}$,
I.~Puljak$^{5}$,
J.~R. Garcia$^{7}$,
I.~Reichardt$^{4}$,\newauthor
W.~Rhode$^{16}$,
M.~Rib\'o$^{20}$,
J.~Rico$^{13}$,
T.~Saito$^{21}$,
K.~Satalecka$^{12}$,
S.~Schroeder$^{16}$,\newauthor
T.~Schweizer$^{7}$,
A.~Sillanp\"a\"a$^{18}$,
J.~Sitarek$^{11}$,
I.~\v{S}nidari\'c$^{5}$,
D.~Sobczynska$^{11}$,\newauthor
A.~Stamerra$^{3}$,
M.~Strzys$^{7}$,
T.~Suri\'c$^{5}$,
L.~Takalo$^{18}$,
F.~Tavecchio$^{3}$,
P.~Temnikov$^{22}$,\newauthor
T.~Terzi\'c$^{5}$,
D.~Tescaro$^{4}$,
M.~Teshima$^{7,21}$,
D.~F.~Torres$^{24}$,
N.~Torres-Alb\`a$^{20}$,
A.~Treves$^{2}$,\newauthor
G.~Vanzo$^{9,10}$,
M.~Vazquez Acosta$^{9,10}$,
I.~Vovk$^{7}$,
J.~E.~Ward$^{13}$,
M.~Will$^{9,10}$,\newauthor
D.~Zari\'c$^{5}$(The MAGIC Collaboration),\newauthor
A.~Loh$^{28}$
and J.~Rodriguez$^{28}$\newauthor
(Affiliations can be found after the references)
}

\date{Accepted XXX. Received YYY; in original form ZZZ}

\pubyear{2017}

\begin{document}
\label{firstpage}
\pagerange{\pageref{firstpage}--\pageref{lastpage}}
\maketitle

\begin{abstract}
The microquasar V404 Cygni underwent a series of outbursts in 2015, June 15-31, during which its flux in hard X-rays (20-40 keV) reached about 40
times the Crab Nebula flux. Because of the exceptional interest of the flaring activity from 
this source, observations at several wavelengths were conducted.
The MAGIC telescopes, triggered by the INTEGRAL alerts, followed-up the flaring
source for several nights during the period June 18-27, for more than 10 hours. 
One hour of observation was conducted simultaneously to a giant 22 GHz radio flare and a hint of signal at GeV energies seen
by {\it Fermi}-LAT. The MAGIC observations did not show significant
emission in any of the analysed time intervals. 
The derived flux upper limit, in the energy range 200--1250 GeV, is
4.8$\times 10^{-12}$ ph cm$^{-2}$ s$^{-1}$. We estimate the gamma-ray opacity during the flaring period, which along with our non-detection, points to an inefficient acceleration in the V404\,Cyg jets if VHE emitter is located further than $1\times 10^{10}$ cm from the compact object.
\end{abstract}




\section{Introduction}
\label{intro}

The microquasar V404 Cygni (V404 Cyg), located  at a parallax distance of 
2.39$\pm 0.14$ kpc \citep{2009ApJ...706L.230M}, is a binary system of an
accreting stellar-mass black hole from a companion star. 
  The black hole mass estimation ranges from about 8 to 15 M$_{\sun}$, 
while the companion star mass is $0.7^{+0.3}_{-0.2}$ M$_{\sun}$ \citep{Casares94, Khargharia10, Shahbaz94}. The system inclination 
angle is 67$^{\circ}$ $^{+3}_{-1}$ \citep{Shahbaz94, Khargharia10}
and the
system orbital period is 6.5 days \citep{Casares94}.
This low-mass X-ray binary (LMXB) showed at least four periods of outbursting activity: the one that led to its discovery in 1989 detected by the Ginga X-ray satellite
\citep{Ginga}, two previous ones in 1938 and 1956 observed in optical and later associated with V404 Cyg \citep{Richter}, and the latest in 2015. 

In June 2015, the system underwent an exceptional 
flaring episode. From the 15th to the end of June the 
bursting activity was registered by several hard X-ray 
satellites, like \textit{Swift} and INTEGRAL \citep{SwiftGCN,INTEGRALAtel}.
It reached a flux about 40 times larger than the Crab Nebula one 
in the 20--40 keV energy band \citep{V404Rodriguez2015}.
The alerts from these instruments triggered follow-up observations from many other instruments from radio \citep{2015ATel_7716_1T, Mooley} to very high
energies \citep{VERITASV404}. 
Recently \cite{2016Natur_531_341S} claimed the detection of the 511 keV gamma signal from electron-positron annihilation in the June V404 Cyg outburst. 
  In agreement with the models, the variability of the annihilation
component suggests that it is produced in the hot 
plasma situated in the inner parts of the accretion 
disk (the so-called corona). 
On the other hand, the possible excess 
seen in the {\it Fermi}-LAT \citep{V404_Fermi}, in 
temporal coincidence with a giant radio 
flare \citep{2015ATel_7716_1T} suggests that the HE 
emission, in the MeV-GeV energy range, originates 
inside the relativistic jet. 
Furthermore the observations of an orphan flare in the near 
Infrared \citep{V404_Tanaka} and the fast variability of the 
optical polarisation \citep{lipunov, shahbaz} indicate the 
presence of a jet. \citet{V404_Tanaka} derive the jet 
parameters, like the magnetic field, and constrain the
emission zone. 

\begin{figure*}
 	\includegraphics[clip=true, width=1.3\columnwidth]{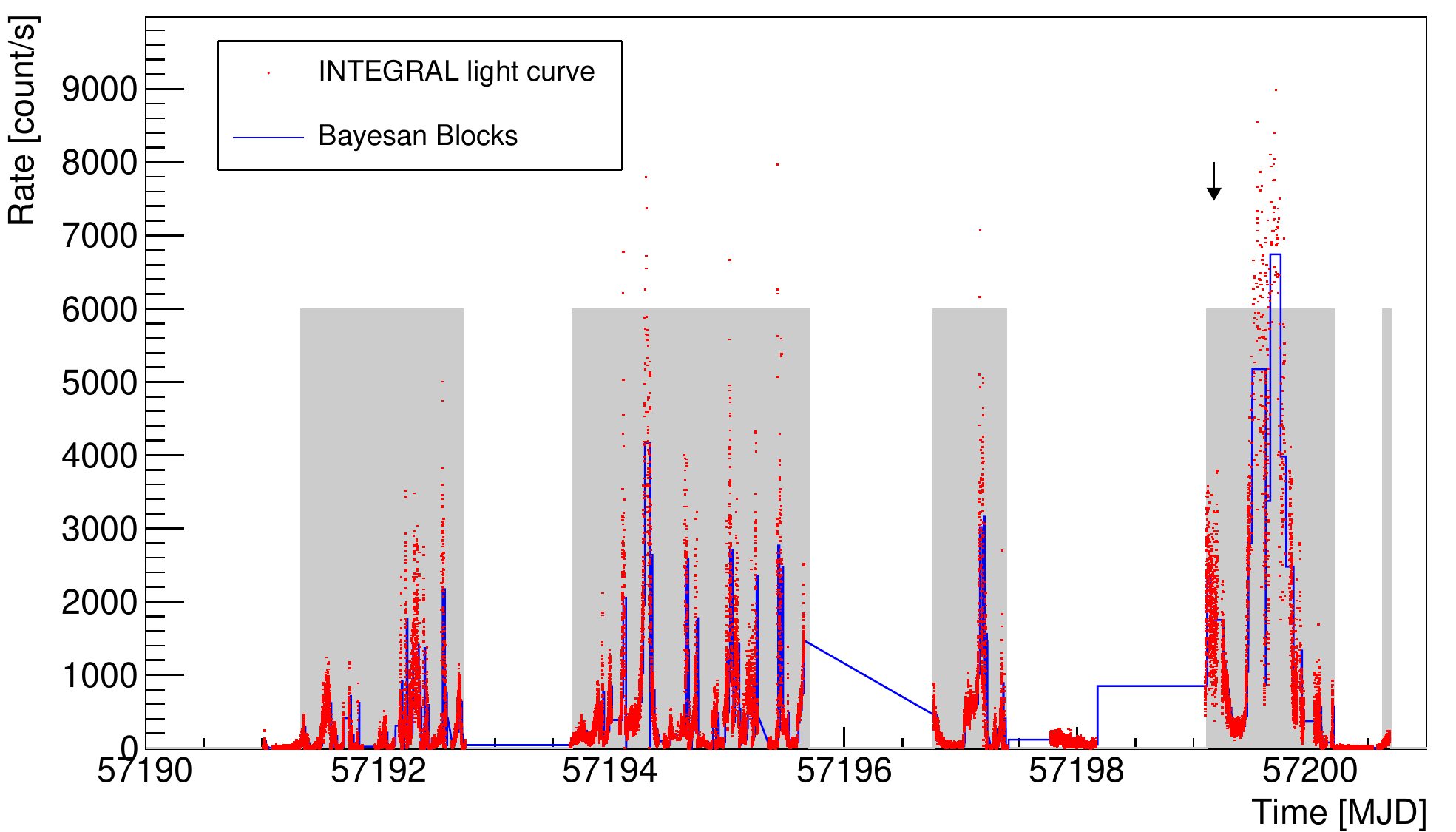}
     \caption{INTEGRAL light curve (red points) in the energy range 20--40 keV with the definition of the flaring interval. The time intervals with the highest flaring activity (gray bands) used in the analysis of MAGIC data are defined following the Bayesian Block method. The arrow refers to the 
     peak of the {\it Fermi}-LAT hint of signal.}
     \label{fig:INTEGRAL_LC}
 \end{figure*}

\begin{figure}
\includegraphics[width=\columnwidth]{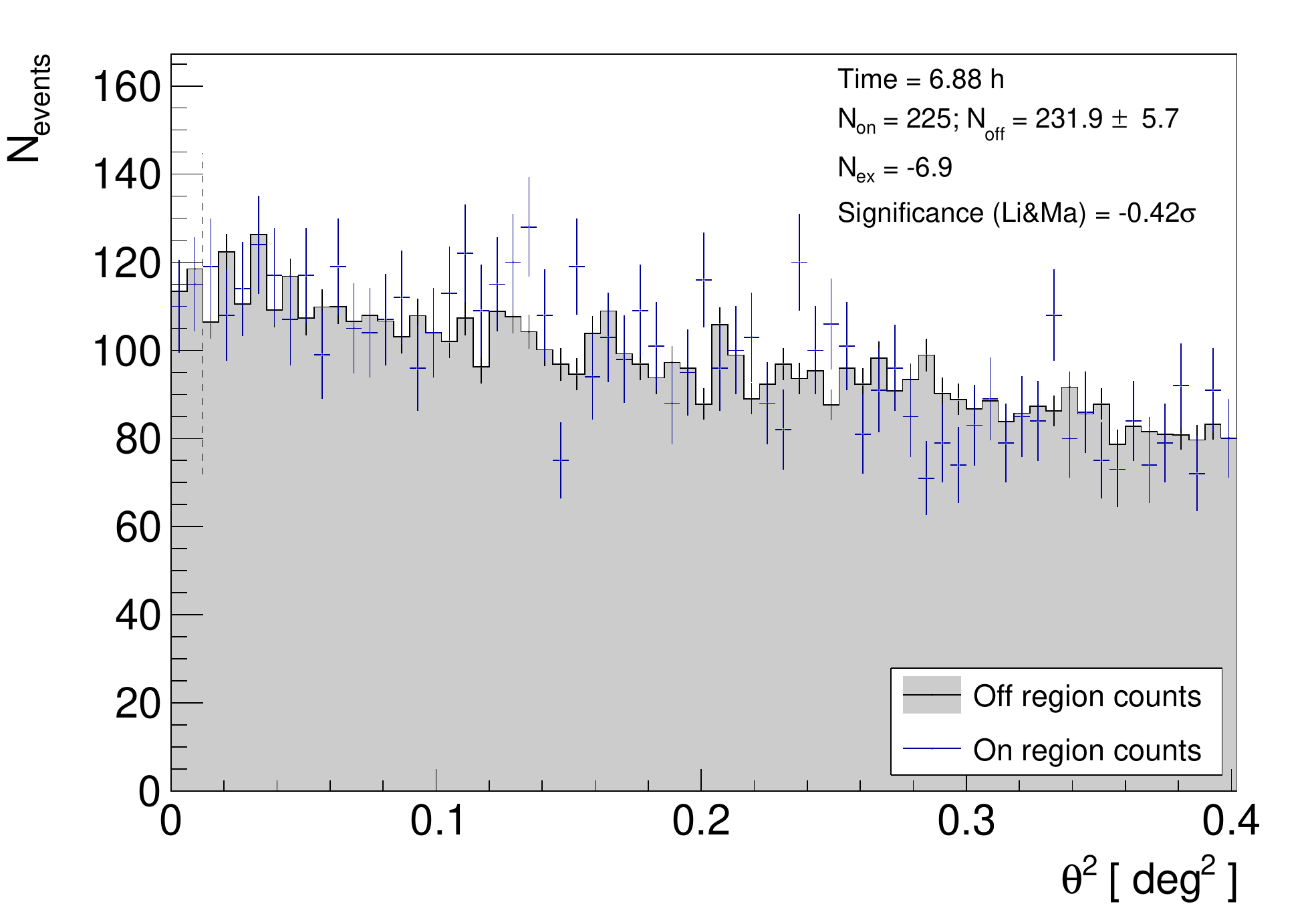}
     \caption{Distribution of the square of the angular distance from the position of the source. In the 7 hours accumulated there is no evidence of signal from V404 Cyg in the MAGIC data.}
     \label{fig:theta2}
 \end{figure}

Very high energy (VHE; E$\gtrsim50$ GeV) gamma-ray 
emission from microquasars has been theoretically predicted in association with the jets where 
relativistic particles are accelerated. VHE radiation could be produced via leptonic \citep[e.g.,][]{2006A&A_447_263B} or hadronic processes \citep[e.g.,][]{2003A&A...410L...1R}. 
IC process on photons from the companion star was 
proposed as the most likely scenario
in the case of two 
microquasars detected in the HE regime: the high-mass X-ray binaries Cygnus\,X-1 \citep{2016arXiv160505914Z, 2016arXiv160705059Z} 
and Cygnus\,X-3 \citep{AgileX3, FermiX3}. In the case
of the possible HE detection of the high-mass X-ray binary SS433 
\citep{Bordas15} the proposed emission mechanism is 
hadronic via proton-proton collisions. 
On the other hand LMXBs, composed of cold and old stars, do not provide a proper photon field target for this process to take place. In LMXBs the dominant processes in the leptonic scenario are syncrotron and 
syncrotron self-Compton emissions from an extended dissipation region in the jet \citep{2015ApJ_806_168Z}.
Differently from HMXBs where the dense matter environments
favours emission from neutral pion decay \citep{Bosch-Ramon2009}, in LMXBs the donor star presents weak winds.
Therefore in the hadronic scenario, photo-pion production
could be considered as the emission mechanism instead
\citep{Levison2001}. In the innermost dissipation region 
of the jet, photon-pions are produced at the $\Delta$ 
resonance by the interaction of accelerated protons and 
external X-ray photons entering the jet. 
Given the lack of targets provided by the low-mass 
companion star in LMXB (like V404 Cyg), gamma rays, 
are expected to be produced inside the relativistic jets 
and in particular where they are most compact, 
like at their base. 
According to models, gamma rays are created by the 
interaction of the particles in the jet with the 
radiation and magnetic fields in the jet itself 
\citep[see e.g.,][]{2006A&A_447_263B,VilaRomero,VieyroRomero}.

Triggered by the INTEGRAL alerts, MAGIC observed V404 Cyg for several nights between June 18th and 27th 2015, 
collecting more than 10 hours. In Section~\ref{obs},  
we present the observations and the instrument 
overview. The analysis of the night wise observations
and the focused analysis following the INTEGRAL light curve 
are presented in Section~\ref{results}. Finally,
we discuss the possible physical implication of 
the results of the MAGIC observations in Section~\ref{discus}.

\begin{table}
	\centering
	\caption{Time intervals selected by the Bayesian Block 
    algorithm. The start and stop times are in MJD.}
	\label{tab:BB_intervals}
	\begin{tabular}{cc} 
		\hline
		Start & Stop\\
		\hline
        \noalign{\vskip 0.08cm}
		57191.337 & 57192.725 \\ 
        57193.665 & 57195.700 \\ 
        57196.765 & 57197.389 \\ 
        57199.116 & 57200.212 \\
        57200.628 & 57200.695 \\
		\hline
	\end{tabular}
\end{table}

\section{Observations \& data analysis}
\label{obs}

MAGIC is a stereoscopic system of two 17m diameter Imaging Atmospheric Cherenkov Telescopes (IACT). It is located at 2200 m a.s.l. in El 
Roque de los Muchachos Observatory, in La Palma, Spain. 
The performance of the telescopes is described in \cite{MAGIC_performance}: the trigger threshold 
is $\sim 50$ GeV below 30$^\circ$ zenith and the integral sensitivity is
$0.66\pm0.03$\% of the Crab Nebula flux above 220 GeV in 50 hours of observations. 

Most of the MAGIC observations were triggered by the INTEGRAL alerts sent via Gamma-ray Coordinate Network (GCN). The first alert was 
received at 00:08:39 UT on the 18th of June.  MAGIC observations continued until the 27th of June when the INTEGRAL alerts ceased. 
On the night between the 22nd and 23rd of June, the observations were not triggered by any alert, but  scheduled a priori according to a multiwavelength campaign on the V404 
Cyg system. The rest of the observations followed a GCN alert processed by the MAGIC 
Gamma-Ray Burst procedure. 
This procedure allows an automatic and fast re-pointing of the telescopes to the burst position in  $\sim 20$ s. Most of the observations were performed
during the strongest hard X-ray flares. In total, MAGIC observed the microquasar for 8 
non-consecutive nights collecting more than 10 hours of data, some coinciding with observations at other energies.

The data were analyzed using the MAGIC software, MARS   
\citep{MARS}, version 2-16-0. Standard event cuts are used to  
improve the signal to background ratio in the MAGIC data as 
described in \cite{MAGIC_performance}. The selections applied 
to estimate the significance of the source are based on 
\textit{hadronness}, $\Theta^2$ 
  and on the size of the shower images.  
The \textit{hadronness} is 
a variable to quantify how likely 
is that a given event was produced by a hadronic atmospheric shower, while the $\Theta$ is the angular distance of each 
event from the position of the source in the camera plane.

\begin{table*}
    \centering
 	\caption{MAGIC observation periods of V404 Cyg. For each
    night the observation interval and duration is reported together with the
    detection significance for that night. In the last   
    column the integral flux upper limits for energies 
    between 200 and 1250 GeV are reported. The last 
    row reports the same quantities for the periods 
    selected with the Bayesian block algorithm.}
 	\label{tab:UL_table}
    \begin{threeparttable}
    \renewcommand{\arraystretch}{1.5}
 	\begin{tabular}{Lcccc} 
 		\hline
 		Observation date & Observation MJD & Effective time & Detection   & Flux UL\\ 
 		(June 2015) & & [h] &  significance & (200<E<1250GeV)\\
  		& &  & [$\sigma$] &  [ph/(cm$^{2}$ s)]\\ 
 		\hline
 		18th & 57191.006--57191.146 & 2.99 & -0.43 & 5.1$\times 10^{-12}$\\
 		19th & 57191.960--57192.055 & 1.9 & -0.6 & 1.00$\times 10^{-11}$\\
 		21st & 57193.997--57194.025 & 0.66 & 1.57 & 4.35$\times 10^{-11}$\\
 		22nd & 57195.021--57195.049 57195.103--57195.134 & 1.33 & 0.09 & 1.67$\times 10^{-11}$\\
		23rd & 57196.003--57196.124 & 2.74 & -0.45 & 3.7$\times 10^{-12}$\\
 		26th & 57199.158--57199.204 & 1.03 & -1.41 & 6.6$\times 10^{-12}$\\
 		27th & 57200.085--57200.115 57200.144--57200.202 & 1.97 & -0.57 & 1.23$\times 10^{-11}$\\
 		Selected & See Table \ref{tab:BB_intervals}
& 6.88 & -0.42 & 4.8$\times 10^{-12}$\\
 		\hline
 	\end{tabular}
    \end{threeparttable}
\end{table*}

\section{Results}
\label{results}

To avoid an iterative search over different time bins, 
we assumed that the TeV flares were simultaneous to 
the X-ray ones. We defined the time intervals  
where we search for signal in the MAGIC data, 
to match those of the flares in the INTEGRAL 
light curve. We analysed the 
INTEGRAL-IBIS data (20--40 keV) publicly available 
with the \texttt{osa} software version 10.2\footnote{http://www.isdc.unige.ch/integral/analysis}, 
obtaining the light curve shown in Figure \ref{fig:INTEGRAL_LC}.

The time selection for the MAGIC analysis was performed running a Bayesian block \citep{2013ApJ...764..167S} analysis on the INTEGRAL light curve (see Figure \ref{fig:INTEGRAL_LC}).
The Bayesian block analysis is meant to identify structures in a time series and to divide these features in adaptive time bins called blocks. To partition the light curve, the algorithm (Jackson et al. 2015) maximises a quantity that describes how well a constant flux represents the data in a given block.
Once the blocks are defined we grouped them into intervals that describe each flaring period.
The analysis did not single out periods with distinctively high levels of source activity.
The Bayesian blocks used to determine the limits of the
periods of activity are listed in Table \ref{tab:BB_intervals}. This 
analysis selected in total about 7 h out of the 10 h observed.  

\begin{figure*}
 	\includegraphics[width=1.5\columnwidth]{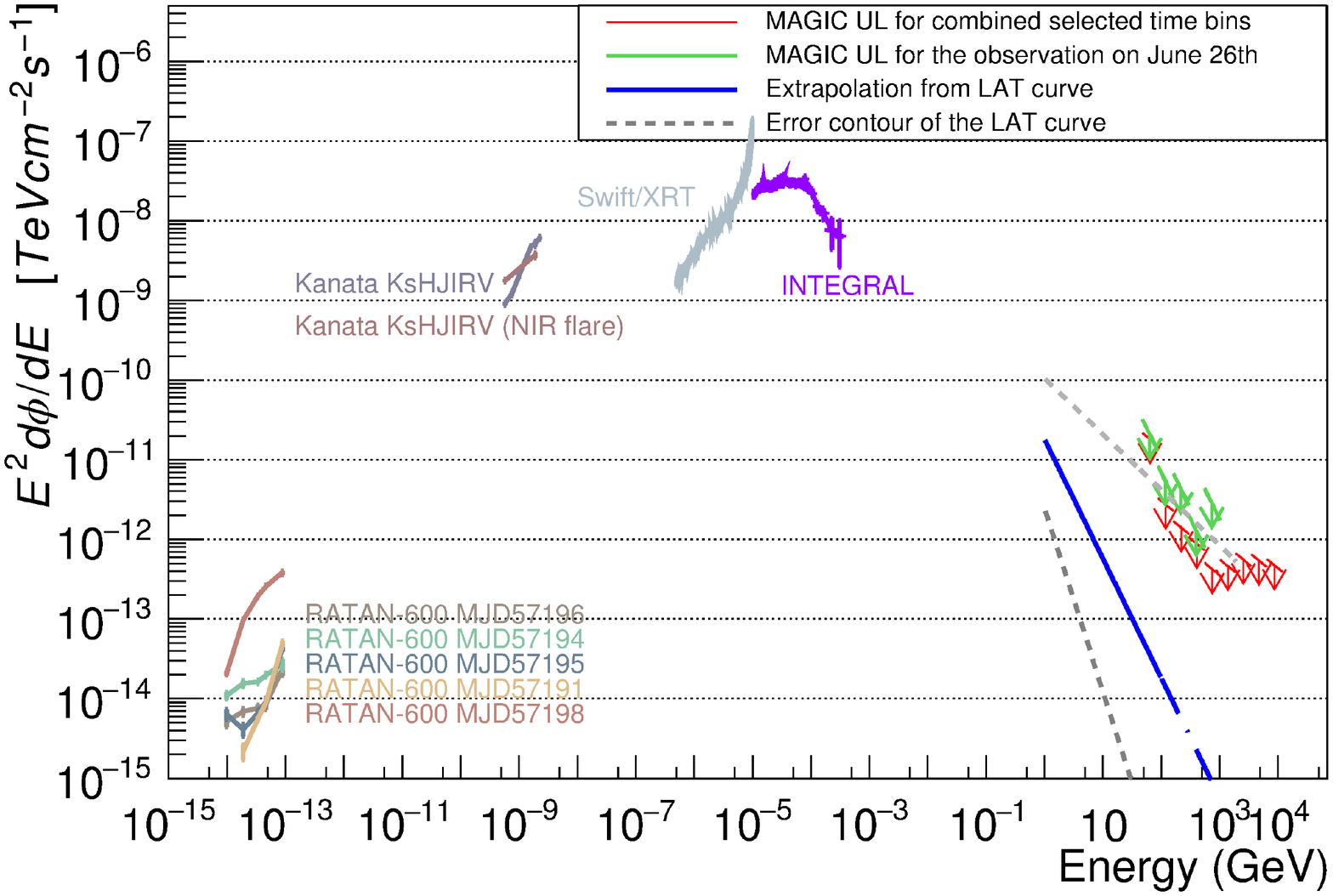}
     \caption{Multiwavelength spectral energy
     distribution of V404 Cyg during the June 
     2015 flaring period. In red, MAGIC ULs are
     given for the combined Bayesian block 
     time bins ($\sim$7 hours) for which a
     power-law function with photon index 
     2.6 was assumed. In green, MAGIC ULs for
     observations on June 26th, simultaneously
     taken with the {\it Fermi}-LAT hint
     \citep{V404_Fermi}. In this case, a photon
     index of 3.5 was applied following 
     {\it Fermi}-LAT results. All the MAGIC upper
     limits are calculated for a 95\% confidence
     level, considering also a 30\% systematic
     uncertainty. The extrapolation 
     of the {\it Fermi}-LAT spectrum is shown in
     blue with 1 $\sigma$ contour (gray dashed
     lines). In the X-ray regime, INTEGRAL 
     \citep[20-40 keV,][]{V404Rodriguez2015} and 
     {\it Swift}-XRT 
     \citep[0.2-10 keV,][]{V404_Tanaka} data are
     depicted. At lower energies, 
     {\it Kanata}-HONIR optical and NIR data are
     shown, taken from \citet{V404_Tanaka}. 
     Finally, RATAN-600 radio data, from
     \citet{2015ATel_7667_1T}, are presented for
     different days during the flaring period.}
     \label{fig:diff_UL}
 \end{figure*}

We searched for VHE gamma-ray emission stacking the MAGIC data of the selected time intervals ($\sim7$ hours).   
We found no significant emission in the $\sim7$  hour
sample (see Figure \ref{fig:theta2}). We found no significant emission also in any of the sub-samples considered (See Table \ref{tab:UL_table}). We then computed 
integral (see Table \ref{tab:UL_table}) and differential upper limits (see Figure \ref{fig:diff_UL}) 
for the observations assuming a power law spectral shape of index -2.6. 
The Li \& Ma \citep{LiMa} method was used to estimate the detection significance while the Rolke method \citep{flux_UL} was used for the computation 
of the upper limits (UL). The upper limits were computed using a Poisson
distribution for the background, requiring a 95\% confidence level and considering a 30\% systematic uncertainty.

%


\cite{V404_Fermi} found in the {\it Fermi}-LAT data evidences for a detection above 4 
$\sigma$ of a source centered 0.65 deg ---which is 
within 95\% of the PSF--- away from V404 Cyg and temporally 
coincident with the brightest radio and hard X-ray flare of this source. The {\it Fermi}-LAT signal is found in the 0.1--100 GeV energy interval
and it peaks at MJD 57199.21$\pm$0.12. MAGIC observation during 
this period starts at MJD 57199.15 and lasts up to MJD 57199.20, 
which is within the 
interval of the {\it Fermi}-LAT excess. For this data 
set we recomputed the differential upper limits using a 
power law with index -3.5 (see green UL in Figure \ref{fig:diff_UL}) according to the LAT analysis presented in \cite{V404_Fermi}. 
The MAGIC upper limits are two order of magnitude higher
than the extrapolation of the {\it Fermi}-LAT spectrum 
(see Figure \ref{fig:diff_UL}). 


\section{Discussion}
\label{discus}

MAGIC observed V404 Cyg for several nights during an outbursting period for a total amount of 
about 10 hours. The analysis of the data
resulted in a non-detection and both differential and integral upper limits have been 
computed. The luminosity upper limits calculated for the full observation period,  
considering the source at a distance of 2.4 kpc, is $\sim 2 \times 10^{33}$ 
erg s$^{-1}$, in contrast with the extreme luminosity emitted in the X-ray band 
($\sim 2 \times 10^{38}$ erg s$^{-1}$, \citealt{V404Rodriguez2015}) and other wavelengths.

The emission of microquasars at VHE is still under 
debate. Processes similar to those taking place in AGN occur
also in microquasars, but at a quite different scale.
Similarly to quasars, microquasars develop 
jets, possibly relativistic, at least in their X-ray hard state \citep{2004MNRAS_355_1105F}. 
If the acceleration that takes place in the jets is efficient enough, VHE photon fluxes could reach  $10^{-13}$ -- $10^{-12}$ ph cm$^{-2}$ s$^{-1}$ (for
an object at about 5 kpc) \citep{2006A&A_447_263B, 2015ApJ_806_168Z, 2015MNRAS_449_34K} making them detectable by this or next generation of IACT.

During the June 2015 outburst of V404 Cyg, there are 
convincing evidences of jet emission given by the
optical observations \citep{V404_Tanaka, lipunov, shahbaz}. 
In particular on the 26th of June, a hint of detection ($\sim4$ $\sigma$) in the {\it Fermi}-LAT data has been reported by \cite{V404_Fermi}. Moreover, the presence 
of a giant radio flare \citep{2015ATel_7716_1T}, an increase of the hardness ratio in the X-ray band \citep{V404_Fermi} and optical 
fast variability \citep{2016MNRAS.459..554G} indicate that the jet environment dramatically changed on that day. 

MAGIC conducted an extensive campaign dedicated to this source, which includes 1 hour of simultaneous observations with the \textit{Fermi}-LAT excess. No signal 
was detected in any of the time intervals considered. We set an energy flux upper limit from a selected dataset of about $\sim 7$ hours of $\sim 2.9 \times 10^{-12}$ erg cm$^{-2}$ s$^{-1}$. The
upper limit is about 2 orders of magnitude smaller than the flux 
released in the 
GeV regime $\sim 4.2 \times 10^{-10}$ erg cm$^{-2}$ s$^{-1}$ \citep{V404_Fermi}. 
\cite{V404_Tanaka} modeled the spectrum of the jet emission in the case of V404 Cyg, obtaining a total radiated flux of
F$_{\rm rad}$= $1.015 \times 10^{-7}$ erg cm$^{-2}$ s$^{-1}$. We compare the flux upper limit obtained from our data with the total radiated flux from this model: the resulting efficiency for VHE gamma emission is lower than 0.003\%.
 
Models predict TeV emission from this type of systems under efficient particle
acceleration on the jets \citep{1999MNRAS.302..253A, 2015ApJ_806_168Z} or strong hadronic jet component
\citep{VilaRomero}. If produced, VHE gamma rays may  annihilate via pair creation in the vicinity of the emitting region. For gamma rays in an energy range 
between 200 GeV -- 1.25 TeV, the largest cross section
occurs with NIR photons.  
For a low-mass microquasar, like V404~Cyg, the contribution 
of the NIR photon field from the companion star (with a 
bolometric luminosity of $\sim10^{32}$ erg s$^{-1}$) is 
very low. During the period of flaring activity, disk and jet contributions are expected to dominate. During the outburst 
activity of June 2015, the magnitude of the K-band 
reached m=10.4 \citep{PANIC_ATel}, leading to a 
luminosity on the NIR regime of $L_{NIR}=\nu 
\phi_{m=0}4\pi d^{2}10^{-m/2.5}=4.1\times 10^{34}$ erg s$^{-1}$, where $\nu$ is the frequency for the 2.2~$\mu$m 
K-band, $\phi_{m=0}=670$ Jy is the K-band reference flux 
and $d=2.4$ kpc is the distance to the source. 
The detected 
NIR radiation from V404~Cyg during this flaring period, 
was expected to be dominated by optically-thick 
synchrotron emission from the jet or to be originated 
inside the accretion flow, given the lack of evidence of 
polarization \citep{V404_Tanaka}. Consequently, stronger 
gamma-ray absorption is expected at the base of the jets. 
The gamma-ray opacity due to NIR radiation inside 
V404~Cyg can be estimated as $\tau_{\gamma\gamma} \sim \sigma_{\gamma \gamma} \cdot n_{NIR} \cdot r$, given by \cite{Aharonian2005}. The 
cross-section of the interaction is defined by $\sigma_{\gamma \gamma}$, whose 
value is $\sim 1\times 10^{-25}$ cm$^{2}$. The NIR photon density is 
calculated as $n_{NIR}=L_{NIR}/\pi r^{2}c\epsilon$: r is the radius of 
the jet where NIR photons are expected to be emitted;
 $c$ is the speed of light and $\epsilon \sim 1\times 10^{-12}$~erg is 
the energy of the target photon field. Assuming the aforementioned
luminosity of $L_{NIR}=4.1\times 10^{34}$~erg~s$^{-1}$, the gamma-ray 
opacity at a typical radius $r\sim 1\times 10^{10}$~cm may be relevant 
enough to avoid VHE emission above 200 GeV. Moreover, if IC on X-rays at 
the base of the jets ($r\lesssim 1\times 10^{10}$ cm) is produced, this 
could already prevent electrons to reach the TeV regime, unless the 
particle acceleration rate in V404\,Cyg is close to the maximum 
achievable including specific magnetic field conditions (see e.g. 
\citealt{Khangulyan2008}). 
On the other hand, VHE photon absorption becomes negligible for
$r>1\times 10^{10}$ 
cm. Thus, if the VHE emission is produced in the same region as HE 
radiation ($r\gtrsim 1\times 10^{11}$ cm, to avoid HE photon absorption in the X-ray photon field), then it would not be significantly affected by pair production
attenuation ($\sigma_{\gamma \gamma}< 1$). Therefore a VHE emitter at $r\gtrsim 1\times 10^{10}$ cm, along to the non-detection by MAGIC,
suggests either a low particle acceleration rate inside the
V404\,Cyg jets or not enough energetics of the VHE emitter.

\section*{Acknowledgements}

We are grateful to Lucia Pavan at UNIGE for the support on the INTEGRAL-IBIS data analysis. 
We would like to thank Dr. Rodriguez and 
Prof. Tanaka for 
providing the multi-wavelength data.
We thank the anonymous reviewer for the useful comments that helped to set our results in a broader multi-wavelength context. 
We would like to thank the Instituto de Astrof\'{\i}sica de Canarias 
for the excellent working conditions at the Observatorio del Roque de los 
Muchachos in La Palma. The financial support of the German BMBF and MPG, 
the Italian INFN and INAF, the Swiss National Fund SNF, the ERDF under 
the Spanish MINECO (FPA2015-69818-P, FPA2012-36668, FPA2015-68378-P, 
FPA2015-69210-C6-2-R, FPA2015-69210-C6-4-R, FPA2015-69210-C6-6-R, 
AYA2015-71042-P, AYA2016-76012-C3-1-P, ESP2015-71662-C2-2-P, 
CSD2009-00064), and the Japanese JSPS and MEXT is gratefully 
acknowledged. This work was also supported by the Spanish Centro de 
Excelencia ``Severo Ochoa'' SEV-2012-0234 and SEV-2015-0548, and Unidad 
de Excelencia ``Mar\'{\i}a de Maeztu'' MDM-2014-0369, by the Croatian 
Science Foundation (HrZZ) Project 09/176 and the University of Rijeka 
Project 13.12.1.3.02, by the DFG Collaborative Research Centers SFB823/C4 
and SFB876/C3, and by the Polish MNiSzW grant 2016/22/M/ST9/00382.







\vspace*{0.5cm}
\noindent
$^{1}$ {ETH Zurich, CH-8093 Zurich, Switzerland} \\
$^{2}$ {Universit\`a di Udine, and INFN Trieste, I-33100 Udine, Italy} \\
$^{3}$ {INAF National Institute for Astrophysics, I-00136 Rome, Italy} \\
$^{4}$ {Universit\`a di Padova and INFN, I-35131 Padova, Italy} \\
$^{5}$ {Croatian MAGIC Consortium, Rudjer Boskovic Institute, University of Rijeka, University of Split - FESB, University of Zagreb - FER, University of Osijek,Croatia} \\
$^{6}$ {Saha Institute of Nuclear Physics, 1/AF Bidhannagar, Salt Lake, Sector-1, Kolkata 700064, India} \\
$^{7}$ {Max-Planck-Institut f\"ur Physik, D-80805 M\"unchen, Germany} \\
$^{8}$ {Universidad Complutense, E-28040 Madrid, Spain} \\
$^{9}$ {Inst. de Astrof\'isica de Canarias, E-38200 La Laguna, Tenerife, Spain} \\
$^{10}$ {Universidad de La Laguna, Dpto. Astrof\'isica, E-38206 La Laguna, Tenerife, Spain} \\
$^{11}$ {University of \L\'od\'z, PL-90236 Lodz, Poland} \\
$^{12}$ {Deutsches Elektronen-Synchrotron (DESY), D-15738 Zeuthen, Germany} \\
$^{13}$ {Institut de Fisica d'Altes Energies (IFAE), The Barcelona Institute of Science and Technology, Campus UAB, 08193 Bellaterra (Barcelona), Spain} \\
$^{14}$ {Universit\`a  di Siena, and INFN Pisa, I-53100 Siena, Italy} \\
$^{15}$ {Institute for Space Sciences (CSIC/IEEC), E-08193 Barcelona, Spain} \\
$^{16}$ {Technische Universit\"at Dortmund, D-44221 Dortmund, Germany} \\
$^{17}$ {Universit\"at W\"urzburg, D-97074 W\"urzburg, Germany} \\
$^{18}$ {Finnish MAGIC Consortium, Tuorla Observatory, University of Turku and Astronomy Division, University of Oulu, Finland} \\
$^{19}$ {Unitat de F\'isica de les Radiacions, Departament de F\'isica, and CERES-IEEC, Universitat Aut\`onoma de Barcelona, E-08193 Bellaterra, Spain} \\
$^{20}$ {Universitat de Barcelona, ICC, IEEC-UB, E-08028 Barcelona, Spain} \\
$^{21}$ {Japanese MAGIC Consortium, ICRR, The University of Tokyo, Department of Physics and Hakubi Center, Kyoto University, Tokai University, The University of Tokushima, Japan} \\
$^{22}$ {Inst. for Nucl. Research and Nucl. Energy, BG-1784 Sofia, Bulgaria} \\
$^{23}$ {Universit\`a di Pisa, and INFN Pisa, I-56126 Pisa, Italy} \\
$^{24}$ {ICREA and Institute for Space Sciences (CSIC/IEEC), E-08193 Barcelona, Spain} \\
$^{25}$ {Humboldt University of Berlin, Institut f\"ur Physik Newtonstr. 15, 12489 Berlin Germany,}\\
$^{26}$ {also at University of Trieste,}\\
$^{27}$ {now at Finnish Centre for Astronomy with ESO (FINCA), Turku, Finland,}\\
$^{28}$ {Laboratoire AIM (CEA/IRFU - CNRS/INSU - University Paris Diderot}\\
%
\bsp	
\label{lastpage}
\end{document}